\documentclass[prl,nofootinbib,twocolumn,superscriptaddress]{revtex4}
\def\mysection#1{{\bf #1.} }
\def\mysections#1{{\bf #1.} }

\newcommand{\be}{\begin{equation}}
\newcommand{\ee}{\end{equation}}
\newcommand{\bea}{\begin{eqnarray}}
\newcommand{\eea}{\end{eqnarray}}
\newcommand{\beq}{\begin{equation}}
\newcommand{\eeq}{\end{equation}}
\def\beqa{\begin{eqnarray}}
\def\eeqa{\end{eqnarray}}
\newcommand{\no}{\nonumber}
\def\lsim{\mathrel{\rlap{\lower4pt\hbox{\hskip1pt$\sim$}}
    \raise1pt\hbox{$<$}}}         
\def\gsim{\mathrel{\rlap{\lower4pt\hbox{\hskip1pt$\sim$}}
    \raise1pt\hbox{$>$}}}         
\def\Mfn{M_{\rm F}}
\def\Sho{S_{H}}
\def\Sle{S_{L}}
\def\Slb{\bar S_{L}}
\def\lamh{\lambda_{H}}
\def\laml{\lambda_{L}}

\begin{document}

{\hspace*{13cm}\vbox{\hbox{WIS/10/04-Apr-DPP}
}}

\vspace*{-10mm}

\title{\boldmath 
The Importance of Being Majorana: \\
Neutrinos versus Charged
Fermions in Flavor Models}

\author{Yosef Nir}\email{yosef.nir@weizmann.ac.il}
\affiliation{Department of Particle Physics,
  Weizmann Institute of Science, Rehovot 76100, Israel}

\author{Yael Shadmi}\email{yshadmi@physics.technion.ac.il}
\affiliation{Physics Department, Technion--Israel Institute of
  Technology, Haifa 32000, Israel}

\vspace*{1cm}

\begin{abstract}
We argue that  
neutrino flavor parameters may exhibit features that
are very different from those of quarks and charged
leptons. Specifically, within the Froggatt-Nielsen (FN) framework, charged
fermion parameters depend on the ratio between two scales, while for
neutrinos a third scale---that of lepton number breaking---is
involved. Consequently, the selection rules for neutrinos may be
different. In particular, if the scale of lepton number breaking is
similar to the scale of horizontal symmetry breaking, neutrinos
may become flavor-blind even if they carry different horizontal charges. 
This provides an attractive mechanism for neutrino flavor anarchy.
\end{abstract}

\maketitle

\mysection{Introduction}
The measured neutrino flavor parameters are neither manifestly small
(apart from the overall mass scale) nor manifestly hierarchical. The
two measured mixing angles are ${\cal O}(1)$ and the measured mass
ratio is ${\cal O}(0.2)$ or larger. With the upper bound on the third mixing
angle of ${\cal O}(0.2)$, and with no information on the remaining mass
ratio and CP violating phases, it could well be that {\it all}
neutrino flavor parameters are non-hierarchical, that is, anarchical
\cite{Hall:1999sn}~(see however~\cite{Altarelli:2002sg}).  
This is in sharp contrast to the charged fermion
flavor parameters. Of these, only two parameters---the top Yukawa and the
KM phase---are  ${\cal O}(1)$, while all other eleven 
parameters---eight masses and three mixing angles---are small and hierarchical.

It is of course possible that yet-unmeasured neutrino parameters
($\theta_{13}$ and/or $m_1/m_2$) are small, and there is hierarchy in all
sectors. We assume here that this is not the case. Then, it is
interesting to understand the reason for the difference between the
flavor structure of neutral and charged fermions. This difference
could be accidental. For example, one could imagine that the flavor
structure is a result of an approximate symmetry, and it just so
happens that all lepton doublets carry the same charge under this
symmetry (see, for example, \cite{Berger:2000sc}). In other words,
each of the sectors---up, down, charged lepton and neutrino---could
equally well be hierarchical or accidentally anarchical. However, a far
more intriguing possibility is that the difference is due to the fact
that, of all the standard model fermions, only neutrinos are Majorana
fermions. Then the measured parameters reflect the interplay between
flavor physics and lepton number violation. It is this interplay that
we wish to explore.  

In order to relate the flavor structure and the
Majorana/Dirac nature of fermions, one must work within a framework
that explains the flavor hierarchy of quarks and charged leptons. 
One of the most attractive such frameworks is the Froggatt-Nielsen
(FN) mechanism \cite{Froggatt:1978nt}. One assumes an Abelian
horizontal symmetry that is broken by a small parameter
near some high ``flavor scale'', $\Mfn$. 
This implies various selection rules for the flavor parameters
of the standard model. We assume that the smallness of the over-all
scale of neutrino masses, is {\it not} a result of the FN selection rules
but rather of the see-saw mechanism \cite{Gell-Mann:vs,yanagida}. 
Neutrino masses are thus universally small because the mass of singlet
Majorana neutrinos or, equivalently, the scale of lepton number
violation, $M_L$, is very high. We will show that
the existence of the scale $M_L$, on top of the FN scale
$\Mfn$, has a crucial impact on neutrino flavor parameters.

\mysection{The Supersymmetric Froggatt-Nielsen Framework}
We consider supersymmetric Froggatt-Nielsen
models~\cite{Leurer:1992wg}.\footnote{Supersymmetry affects our study
  in three ways:
  (i) Superpotential flavor parameters are governed by holomorphy 
in addition to
  the horizontal symmetry;
  (ii) The Higgs sector consists of two doublets; 
  (iii) Both the see-saw and FN mechanisms introduce new heavy
  particles with Yukawa couplings to the Higgs field. In the absence
  of supersymmetry, severe fine-tuning problems would arise.}
We assume the following symmetries:
\beq\label{symm}
G_{\rm SM} \times U(1)_H \times U(1)_L\ .
\eeq
Here $G_{\rm SM}$ is the SM gauge group, spontaneously broken by two
Higgs doublets, $\phi_u(1,2)_{+1/2}$ and
$\phi_d(1,2)_{-1/2}$. Supersymmetry is softly broken, but since its
breaking is irrelevant to our investigation, we do not specify the
breaking mechanism here. The $U(1)_H$ factor is the horizontal symmetry,
which we take to be a $U(1)$ for simplicity. To avoid the issue of
global symmetry breaking by strong gravity effects, as well as
Goldstone bosons, we could choose the horizontal symmetry to be a
(gauged) discrete symmetry. We assume that it is broken by the VEV of
a {\it single} scalar field $\Sho$ (more accurately, $\Sho$ is the scalar
component in a chiral supermultiplet) that is a singlet of $G_{\rm
SM}\times U(1)_L$ and carries charge $-1$ under $U(1)_H$. This choice
just sets the overall normalization of $H$-charges. The $U(1)_L$
symmetry is lepton number. We assume that it is broken by the VEVs of
two scalar fields $\Sle$ and $\Slb$ that are singlets of $G_{\rm
  SM}\times U(1)_H$ and carry charges $+2$ and $-2$, respectively,
under $U(1)_L$. 
The two VEVs are equal in magnitude.\footnote{A
  single $\langle\Sho\rangle$ would arise naturally for a
  pseudo-anomalous $U(1)_H$~\cite{Ibanez:ig}. 
  Equal VEVs
  $\langle\Sle\rangle=\langle\Slb\rangle$ would be necessary to preserve
  supersymmetry with a gauged, non-anomalous $U(1)_L$. 
  Similarly, two spurions of equal VEVs and opposite $U(1)_H$ charges
  would be necessary if $U(1)_H$ is a gauged, non-anomalous 
  symmetry~\cite{Nir:1999xp}.
  Neither ingredient is important for our purposes.}

The symmetries~(\ref{symm}) forbid neutrino masses, and, for
appropriate choices of the quark and lepton horizontal charges, most
of the charged fermion masses. These masses and couplings are
generated however when integrating out new heavy fields. These heavy
``FN fields'' have charges similar to those of the SM quarks and
leptons (that is, $\pm2/3$, $\mp1/3$, $\mp1$ and $0$), but appear in
vector representations of $G_{\rm SM}\times U(1)_H$. If the FN fields
are vector-like also under $U(1)_L$---as is always the case for the
charged fields---they have masses at a high scale $\Mfn$ (possibly the
Planck scale). Heavy singlet neutrinos may, however, be chiral under
$U(1)_L$. In that case, they acquire masses at the scale of lepton
number breaking, $M_L\lsim\Mfn$.   
Thus there are four relevant mass scales in our framework:
\begin{enumerate}
\item $\langle\phi_{u,d}\rangle$, the electroweak breaking scale;
\item $M_L\equiv\langle\Sle\rangle=\langle\Slb\rangle$, the lepton
  number breaking scale; 
\item $M_H\equiv\langle \Sho\rangle$, the horizontal symmetry breaking scale;
\item $\Mfn$, the mass scale of Froggatt-Nielsen
vector-like quarks and leptons.  
\end{enumerate}
We assume the following hierarchies:
\beq\label{scahie}
\langle\phi_{u,d}\rangle\ll M_L,\ M_H,\ \Mfn ,\ \ \ \ \ 
M_L\lsim \Mfn,
\eeq
\beq\label{deflam}
\lamh\equiv\frac{M_H}{\Mfn}\ll1\ .
\eeq
For concreteness, we often use $\lamh\sim0.2$, inspired
by the value of the Cabibbo angle which one may attempt to explain as
being suppressed by a single power of the ratio $\langle
\Sho\rangle/\Mfn$. The precise numerical value is, however, irrelevant
for our conclusions.

Note that we do not specify the relative sizes of the lepton-number
breaking scale, $M_L$, and the horizontal symmetry
breaking scale $M_H$. In the following, we will explore
the impact of different hierarchies between these scales on
neutrino parameters.

\mysection{Charged Fermion Parameters}
To understand the resulting quark flavor structure, it is sufficient to
consider a low energy effective theory that includes only the MSSM
fields. The theory has a $U(1)_H$ symmetry which is {\it explicitly}
broken by the spurion $\lamh\sim0.2$ of $U(1)_H$-charge $-1$.  
This leads to the following selection rules:
   \begin{enumerate}
    \item Superpotential terms of integer $H$-charge
      $n\geq0$ are suppressed by $\lamh^n$.
    \item Superpotential terms of negative or non-integer
      $H$-charge vanish.
    \end{enumerate}

These selection rules are sufficient in order to find the parametric
suppression (that is, the $\lamh$ dependence) of the flavor
parameters. In particular, if holomorphic zeros play no role, the
mixing angles and mass ratios are (with $i<j;\ q=u,d$): 
\beq\label{phypar}
V_{ij}\sim\lamh^{H(Q_i)-H(Q_j)},\ \ \
m_i/m_j\sim\lamh^{H(Q_i)-H(Q_j)+H(\bar q_i)-H(\bar q_j)}\ .
\eeq
For example, quark parameters are often accounted for by
the following set of $H$-charges:
\beqa\label{staqua}
&\phi_u(0),\ \phi_d(0),\ \ \ &Q_1(3),\ Q_2(2),\ Q_3(0),\no\\
&\bar u_1(5),\ \bar u_2(2),\ \bar u_3(0),\
\ &\bar d_1(3),\ \bar d_2(2),\ \bar d_3(2),
\eeqa
which imply
\beqa\label{quphpa}
V_{us}\sim\lamh,&\ \ V_{cb}\sim\lamh^2,\ \
&V_{ub}\sim\lamh^3,\no\\
m_u/m_c\sim\lamh^4,&\ \ m_c/m_t\sim\lamh^4,\ \
&m_t/\langle\phi_u\rangle\sim1,\no\\
m_d/m_s\sim\lamh^2,&\ \ m_s/m_b\sim\lamh^2,\ \
&m_b/\langle\phi_d\rangle\sim\lamh^2,
\eeqa
consistent (for $\tan\beta\equiv\langle\phi_u\rangle/
\langle\phi_d\rangle\sim1$)  with the experimental values. 

Let us see how this low energy effective theory arises in a full
high energy FN model. As an example, we focus on the $(c,t)$
sector. We add the following FN fields: 
\beq\label{exfnfi}
\bar U_{-2}+U_{+2},\ \ \bar U_{-1}+U_{+1},\ \ 
\bar U_0+U_0,\ \ \bar U_{+1}+U_{-1}\ .
\eeq
Here $U_h(\bar U_h)$ is an $SU(2)$-singlet quark (antiquark) of
horizontal charge $h$. The mass matrix for rows corresponding to
$(Q_2,Q_3,U_{+2},U_{+1},U_0,U_{-1})$ 
and columns to $(\bar u_2,\bar u_3,\bar U_{-2},\bar U_{-1},\bar
U_0,\bar U_{+1})$ is given by (up to ${\cal O}(1)$-coefficients)
\beq\label{fnmup}
\pmatrix{0&0&\phi_u&0&0&0\cr 0&\phi_u&0&0&0&0\cr
  0&0&\Mfn&\Sho&0&0\cr 0&S&0&\Mfn&\Sho&0\cr
  0&0&0&0&\Mfn&\Sho\cr \Sho&0&0&0&0&\Mfn\cr}\ .
\eeq
When the four heavy FN fields with masses of ${\cal O}(\Mfn)$ are
integrated out, we obtain
\beq
M_u^{(c,t)}\sim\langle\phi_u\rangle\pmatrix{
  \lamh^4&\lamh^2\cr \lamh^2&1}\ ,
\eeq
consistent with $m_c/m_t\sim\lamh^4$ and $|V_{cb}|\sim\lamh^2$.
  
\mysection{Neutrino Parameters}
We assume that neutrino masses arise from the see-saw mechanism, that
is superpotential terms of the form
\beq\label{sesate}
\frac{Z_{ij}}{M_L}\phi_u\phi_u L_iL_j\ .
\eeq
$Z$ is a $3\times3$ matrix of dimensionless Yukawa couplings. We aim
to find the selection rules that apply to it and see if they are
different in a fundamental way from those of charged fermions.
Indeed, even the most naive selection rules \cite{Grossman:1995hk}
have two special features:
\begin{enumerate}
\item The matrix is symmetric, $Z_{ij}=Z_{ji}$. Thus, in contrast to
  the charged fermion case, pairs of entries are related, and we can
  get a (quasi-)degeneracy.
  \item Terms in (\ref{sesate}) that carry a negative $H$ charge,
    $n<0$, might be {\it enhanced} by $\lamh^n$ rather than vanish.
  \end{enumerate}

The lepton number breaking parameters have, however, an even more
profound effect on the selection rules. Specifically, they introduce
an additional parameter, on top of $\lamh$ of eq. (\ref{deflam}),
that breaks the horizontal symmetry and conserves lepton number:
\beq\label{deflamr}
\laml^2\equiv\frac{\langle
  \Sho\rangle^2}{\langle\Sle\rangle\langle\Slb\rangle}=\frac{M_H^2}{M_L^2}\ . 
\eeq
The crucial point is that $\lamh^2/\laml^2$ is neutral under all the
symmetries (\ref{symm}) and therefore can affect the physical
observables in a way that depends sensitively on the details of the full 
high-energy theory. 
Furthermore, the numerical value of $\laml$ depends on the
hierarchy of scales $M_H$ and $M_L$. We can have $\laml\sim\lamh$ or
$\laml>\lamh$ and even $\laml\gsim1$.

Only in the special case that $\laml\sim\lamh$, that is,
$M_L\sim\Mfn$, we expect that neutrinos will have a flavor
hierarchy that is related to the one in the charged fermion
sectors. Generically, however, the structure of the neutrino flavor
parameters depends, in addition to $\lamh$, on $\laml$, and can
be very different from that of quarks and charged lepton masses.
In the next section we give several examples that demonstrate these
statements. 

\mysection{Explicit examples}
We consider a simplified framework of two light active neutrinos. As
an explicit example, we take the two lepton doublets to be $L_{+2}$
and $L_0$, where the sub-index denotes the $H$-charge. We present three
different full high-energy models. The various models exhibit
several interesting features that may arise in the neutrino sector and
demonstrate the sensitivity of  low-energy observables to the full
high-energy theory.

Each of the models is defined by a set of $G_{\rm SM}$-singlet
fields. 
To obtain the light-neutrino mass matrix, we start from
the full renormalizable superpotential allowed by the symmetries.
As above, we omit dimensionless ${\cal O}(1)$ coefficients and, for the
light-neutrino mass matrices, contributions that are subleading
in $\lambda_H/\lambda_L$.
Leptons (antileptons) of $H$-charge $h$ are denoted by
$N_h$ ($\bar N_h$). 

{\bf Model I} has the following (anti)lepton fields:
\beq\label{modela}
L_{+2},\ L_0,\  \bar{N}_{-2},\ N_{+2},\ \bar N_{-1},\ N_{+1},\ 
\bar N_0,\ \bar N_0\ .
\eeq
The mass matrix in this basis is:
\beq
\pmatrix{0&0&\phi_u&0&0&0&0&0\cr 0&0&0&0&0&0&\phi_u&\phi_u\cr
\phi_u&0&0&\Mfn&0&0&0&0\cr 0&0&\Mfn&0&\Sho&0&0&0\cr
0&0&0&\Sho&0&\Mfn&0&0\cr 0&0&0&0&\Mfn&0&\Sho&\Sho\cr
0&\phi_u&0&0&0&\Sho&\Slb&0\cr 0&\phi_u&0&0&0&\Sho&0&\Slb\cr}.
\eeq
The light neutrino mass matrix is given by 
\beq
M_l \sim \frac{\langle\phi_u\rangle^2}{M_L}\pmatrix{
  \lamh^4 & \lamh^2\cr \lamh^2&1\cr}.
\eeq
It leads to the `naive' flavor structure, namely the flavor structure
that would follow if the selection rules were similar to
those of charged fermions \cite{Grossman:1995hk,Binetruy:1996xk}:
\beq\label{naive}
m_1/m_2\sim\lamh^4,\ \ \ \sin\theta\sim\lamh^2\ .
\eeq

{\bf Model II} has the following (anti)lepton fields:
\beq\label{modelb}
L_{+2},\ L_0,\  \bar{N}_{-2},\ N_{+2},\ \bar N_{-1},\ \bar N_{+1},\ 
N_0,\ \bar N_0\ .
\eeq
The mass matrix in this basis is:
\beq
\pmatrix{0&0&\phi_u&0&0&0&0&0\cr 0&0&0&0&0&0&0&\phi_u\cr
\phi_u&0&0&\Mfn&0&0&0&0\cr 0&0&\Mfn&0&\Sho&0&0&0\cr
0&0&0&\Sho&0&\Slb&0&0\cr 0&0&0&0&\Slb&0&\Sho&0\cr
0&0&0&0&0&\Sho&\Sle&\Mfn\cr 0&\phi_u&0&0&0&0&\Mfn&\Slb\cr}.
\eeq
The light neutrino mass matrix is given by 
\beq
M_l \sim \frac{\langle\phi_u\rangle^2}{M_L}\frac{\lamh^2}{\laml^2}\pmatrix{
  \lamh^2\laml^2& \laml^2\cr \laml^2 & 1\cr}.
\eeq
This mass matrix has interesting features:
\begin{enumerate}
  \item  For $\laml^2\sim\lamh^2$, the mixing and hierarchy assume their
    naive values, as in (\ref{naive}).
  \item For $\lamh^2<\laml^2<1$, the mixing is larger,
    $\sin\theta\sim\lambda_L^2$, and the 
    hierarchy is weaker, $m_1/m_2\sim\lambda_L^4$, than the naive
    estimates. 
      \item For $\laml^2>1$, we have a pseudo-Dirac state.
  \end{enumerate}
Since the two spurions, $\lambda_H$ and
$\lambda_L$, appear in the light-neutrino mass matrix, 
the naive selection rules do not necessarily apply,
and a flavor structure unique to
neutrinos, such as a pseudo-Dirac state, may arise.

{\bf Model III} has the following (anti)lepton fields:
\beq\label{modelc}
L_{+2},\ L_0,\  \bar{N}_{-2},\ \bar N_{+2},\ N_{-1},\ N_{+1},\ \bar
N_0,\ \bar N_0\ .
\eeq
The mass matrix in this basis is:
\beq
\pmatrix{0&0&\phi_u&0&0&0&0&0\cr 0&0&0&0&0&0&\phi_u&\phi_u\cr
\phi_u&0&0&\Slb&0&0&0&0\cr 0&0&\Slb&0&\Sho&0&0&0\cr
0&0&0&\Sho&0&\Sle&0&0\cr 0&0&0&0&\Sle&0&\Sho&\Sho\cr
0&\phi_u&0&0&0&\Sho&\Slb&0\cr 0&\phi_u&0&0&0&\Sho&0&\Slb\cr}.
\eeq
The light neutrino mass matrix is given by
\beq
M_l \sim \frac{\phi_u^2}{M_L}\pmatrix{
  \laml^4& \laml^2\cr \laml^2 & 1\cr}\ ,
\eeq
so that
\beq
m_1/m_2\sim\laml^4,\ \ \ \sin\theta\sim\laml^2\ .
\eeq
The following two ranges for $\lambda_L$ are particularly interesting:
\begin{enumerate}
\item For $\laml>1$, we obtain inverted hierarchy: the state with the
  highest FN-charge is the heaviest, in contrast to charged fermions.
\item For $\laml\sim1$, there is no hierarchy in the masses
and mixing angle, {\it i.e.} we have neutrino flavor anarchy. 
\end{enumerate}
We learn that if $U(1)_H$ and $U(1)_L$ are broken at the same scale,
it is quite possible that neutrinos will have no special flavor
structure, even if they come from lepton doublets that carry different
$H$-charges.

\mysection{Conclusions}
The special flavor features (smallness and hierarchy) of quark and
charged lepton masses and CKM mixing angles can be explained by a
spontaneously broken horizontal symmetry, if the breaking scale is
lower than the scale where the breaking is communicated to the light
quarks and leptons.

If the light active neutrinos are Majorana particles and derive their
masses through a seesaw mechanism, an additional scale plays a role in
their flavor structure, that is the scale of lepton number breaking
or, equivalently, the Majorana mass scale of the heavy singlet
neutrinos. This fact may have significant effects on the neutrino
sector. Its flavor parameters may have a hierarchy that is very
different from the charged fermions. Intriguing features, such as
inverted hierarchy or a pseudo-Dirac state, can appear in the neutrino
sector.

In particular,
the neutrino flavor parameters may have no
special structure at all. 
While there is no inherent motivation for neutrino anarchy
in the framework that we investigated, 
it does arise naturally if the horizontal symmetry and
the lepton number symmetry are broken at the same scale.

Thus, if future measurements of neutrino parameters strengthen the
case for flavor anarchy ($|U_{e3}|$ close to the present upper bound
and no quasi-degeneracy among the masses), models that relate the two
scales will be favored.

The ideas presented in this work can be extended in a straightforward
way to realistic, three generation models. It would also be
interesting to explore whether, in other mechanisms that explain the
hierarchy in the charged fermion parameters, the Majorana nature of
neutrinos introduces significant modifications that are particular to
this sector.

\mysections{Acknowledgments}
We thank Srubabati Goswami and D.~Indumathi for useful and enjoyable
discussions. We thank the organizers of the Eighth Workshop on High
Energy Physics Phenomenology (WHEPP-8), Mumbai, where this project
was initiated.  This project was supported by
the Albert Einstein Minerva Center for Theoretical Physics. 
The research of Y.N. is supported by the Israel
Science Foundation founded by the Israel Academy of Sciences and
Humanities, by a Grant from the G.I.F., the German--Israeli
Foundation for Scientific Research and Development, by a grant
from the United States-Israel Binational Science Foundation (BSF),
Jerusalem, Israel, and by EEC RTN contract HPRN-CT-00292-2002. 
The research of Y.S. is supported in part by the Israel Science 
Foundation (ISF) under grant 29/03, 
and by the United States-Israel Binational Science 
Foundation (BSF) under grant 2002020.


\begin{thebibliography}{01}
  \vspace*{3mm}
  
\bibitem{Hall:1999sn}
L.~J.~Hall, H.~Murayama and N.~Weiner,
Phys.\ Rev.\ Lett.\  {\bf 84}, 2572 (2000)
[arXiv:hep-ph/9911341].

\bibitem{Altarelli:2002sg}
G.~Altarelli, F.~Feruglio and I.~Masina,
JHEP {\bf 0301}, 035 (2003)
[arXiv:hep-ph/0210342].

\bibitem{Berger:2000sc}
M.~S.~Berger and K.~Siyeon,
Phys.\ Rev.\ D {\bf 63}, 057302 (2001)
[arXiv:hep-ph/0010245].

\bibitem{Froggatt:1978nt}
C.~D.~Froggatt and H.~B.~Nielsen,
Nucl.\ Phys.\ B {\bf 147}, 277 (1979).

\bibitem{Gell-Mann:vs}
M.~Gell-Mann, P.~Ramond and R.~Slansky,
Print-80-0576 (CERN).

\bibitem{yanagida}
T. Yanagida, in {\it Proc. of Workshop on Unified Theory and Baryon
  Number in the Universe}, eds. O. Sawada and A. Sugamoto (KEK, 1979).

\bibitem{Leurer:1992wg}
M.~Leurer, Y.~Nir and N.~Seiberg,
Nucl.\ Phys.\ B {\bf 398}, 319 (1993)
[arXiv:hep-ph/9212278];
Nucl.\ Phys.\ B {\bf 420}, 468 (1994)
[arXiv:hep-ph/9310320].



\bibitem{Ibanez:ig}
L.~E.~Ibanez and G.~G.~Ross,
Phys.\ Lett.\ B {\bf 332}, 100 (1994)
[arXiv:hep-ph/9403338].

\bibitem{Nir:1999xp}
Y.~Nir and Y.~Shadmi,
JHEP {\bf 9905}, 023 (1999)
[arXiv:hep-ph/9902293].


\bibitem{Grossman:1995hk}
Y.~Grossman and Y.~Nir,
Nucl.\ Phys.\ B {\bf 448}, 30 (1995)
[arXiv:hep-ph/9502418].

\bibitem{Binetruy:1996xk}
P.~Binetruy, S.~Lavignac and P.~Ramond,
Nucl.\ Phys.\ B {\bf 477}, 353 (1996)
[arXiv:hep-ph/9601243].




\end{thebibliography}
\end{document}